\documentclass[a4paper,twoside]{article}
\newcommand{\affil}[1]{$^{\rm #1}$}
\usepackage[authoryear]{natbib}
\bibpunct{(}{)}{;}{a}{}{,}
\usepackage{graphicx}
\date{} 
\newcommand{\ncap}{(n,$\gamma$)}
\title{\large\bf\flushleft Opportunities for Nuclear Astrophysics at FRANZ}
\author{\parbox{\textwidth}{\flushleft
\vspace{-0.5cm}
{\it R.~Reifarth~\affil{A,B,*}, 
L.P.~Chau~\affil{B},
M.~Heil~\affil{A}, 
F.~K{\"a}ppeler~\affil{C}, 
O.~Meusel~\affil{B},
R.~Plag~\affil{A,B}
U.~Ratzinger~\affil{B},
A.~Schempp~\affil{B},
K.~Volk~\affil{B},
   }\\
\vspace{0.4cm}
{\small \affil{A}\,Gesellschaft f{\"u}r Schwerionenforschung mbH, Darmstadt, D-64291, Germany}\\
{\small \affil{B}\,J.W. Goethe-Universit{\"a}t, Frankfurt a.M, D-60438, Germany}\\
{\small \affil{C}\,Forschungszentrum Karlsruhe, Karlsruhe, D-76021, Germany}\\
{\small \affil{*}\,Email: r.reifarth@gsi.de}}}
\begin{document}
\twocolumn[
\begin{changemargin}{.8cm}{.5cm}
\begin{minipage}{.9\textwidth}
\vspace{-1cm}
\maketitle
%
%
\small{\bf Abstract:}

The "Frankfurter Neutronenquelle am Stern-Gerlach-Zentrum" (FRANZ), which is 
currently under development, will be the strongest neutron source in the 
astrophysically interesting energy region in the world. It will be about three 
orders of magnitude more intense than the well-established neutron source at 
the Research Center Karlsruhe (FZK). 

\medskip
\medskip
\end{minipage}
\end{changemargin}
]
\small

\section{Introduction}
\label{ch_intro}

\begin{figure}
  \includegraphics[width=.45\textwidth]{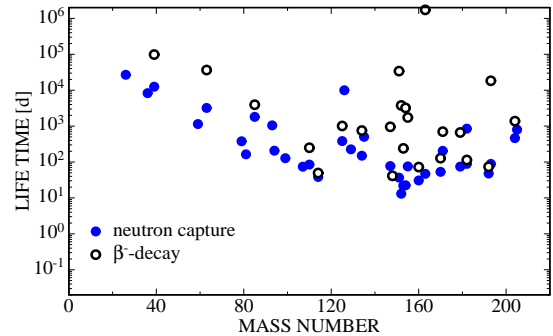}
  \caption{Neutron capture live times (filled circles) for a neutron density of 5$\cdot$10$^8$~cm$^{-3}$ 
  and $\beta^-$ live times (open circles) for unstable isotopes on the s-process path as 
  a function of mass number. Shown are only isotopes where the neutron capture is faster than the
  (terrestrial) $\beta^-$ decay. The neutron capture cross sections are taken from \cite{BBK00}.
  \label{reifarthr_f1}}
\end{figure}

About 50\% of the element abundances beyond iron are produced via slow neutron capture 
nucleosynthesis ($s$ process). Starting at iron-peak seed, the $s$-process mass flow 
follows the neutron rich side of the valley of stability. If different reaction rates 
are comparable, the $s$-process path branches and the branching ratio reflects the 
physical conditions in the interior of the star. Such nuclei are most interesting, 
because they provide the tools to effectively constrain modern models of the stars 
where the nucleosynthesis occurs. As soon as the $\beta^-$ decay is faster than the 
typically competing neutron capture, no branching will take place. Therefore 
experimental neutron capture data for the $s$ process are only needed, if the respective 
neutron capture time under stellar conditions is similar or smaller than the $\beta^-$ 
decay time, which includes all stable isotopes. Depending on the actual neutron density 
during the $s$ process, the "line of interest" is closer to or farther away from the 
valley of stability. 
In a recent estimate the neutron density within the classical $s$-process model 
\cite{KBW89} was estimated to be $n_n~=~(4.94^{+0.60}_{-0.50})\times 10^8~{\rm cm^{-3}}$ 
\cite{RAH03}. Figure~\ref{reifarthr_f1} shows a summary of the neutron capture and 
$\beta^-$ decay times for radioactive isotopes on the neutron rich side of the 
valley of stability, under the condition that the classical neutron capture occurs 
than the terrestrial $\beta^-$ decay. Obviously the vast majority of isotopes, 
where an experimental neutron capture cross section is desirable, have 
$\beta^-$ half-lives of at least hundreds of days.

\begin{figure}
  \includegraphics[width=.45\textwidth]{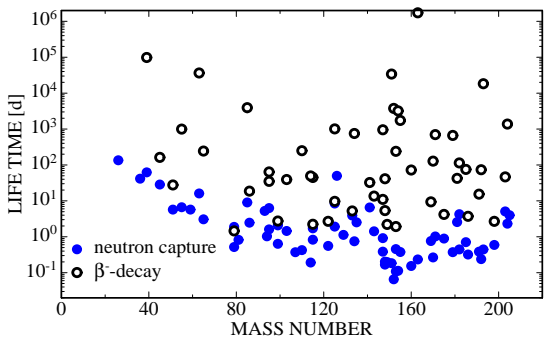}
  \caption{Neutron capture live times (filled circles) for a neutron of density 10$^{11}$~cm$^{-3}$ 
  and $\beta^-$ live times (open circles) for unstable isotopes on the s-process path as 
  a function of mass number. Shown 
  are only isotopes where the neutron capture is faster than the terrestrial $\beta^-$ decay. The neutron 
  capture cross sections are taken from \cite{BBK00}.
  \label{reifarthr_f2}}
\end{figure}

The modern picture of the main $s$-process component refers to the He shell burning phase 
in AGB stars \cite{LHL03}. The highest neutron densities in this model occur during the 
$^{22}$Ne($\alpha$,n) phase and are up to 10$^{11}$ cm$^{-3}$. Figure~\ref{reifarthr_f2} 
shows the same as Figure~\ref{reifarthr_f1}, but for the higher neutron density. Now 
isotopes with half-lives down to a few days can be of interest for the $s$ process reaction 
network.

Improved experimental techniques, especially as far as the neutron source and sample preparation 
are concerned, are necessary to perform direct neutron capture measurements on such isotopes \cite{RHH04}.

In section~\ref{ch_franz} a new approach currently realized at the Goethe University Frankfurt, Germany is 
described. The FRANZ facility will allow energy-dependent neutron capture cross section and activation
experiments in the astrophysically 
interesting energy region with significantly higher neutron fluxes then currently available elsewhere 
\cite{Koe01}.    

In section~\ref{ch_franz_tof} the time-of-flight (TOF) method is described and a comparison
of the new facility with existing facilities is given. 
In section~\ref{ch_franz_activation} the activation method is described and the opportunities and challenges 
at the new facility are discussed. In this approach the limited neutron flux
is overcome by an extremely short distance between the sample and the neutron production target. The trade-off 
is then the very limited information about the energy-dependence of the measured cross section.   

\section{FRANZ}
\label{ch_franz}

As already discussed in section~\ref{ch_intro}, it is desirable to 
improve the currently available
experimental possibilities for neutron capture experiments.  
Spallation or photo-neutron sources require large accelerators, but a small 
accelerator as used for the recent $^{60}$Fe activation at FZK \cite{URS09} 
is best suited for neutron experiments in a university environment. 
This solution has the additional advantage that the neutron spectrum can be 
tailored to the specific energy range of interest. 
 
Among the different options for producing neutrons in the keV region, the 
$^7$Li(p,n)$^7$Be reaction with a threshold of 1.881~MeV is by far the most prolific. 
Near the threshold one can also take advantage of the fact that kinematically 
collimated neutrons can be produced in the energy range up to 100~keV. 
Our approach is therefore to use the existing experience with this method to 
produce neutrons by upgrading the proton source as well as high 
current lithium targets. The last setup at FZK for ToF measurements had a flight 
path of about 80~cm and about 10$^4$ neutrons/s/cm$^2$ at the sample position with  
proton currents of $\approx~2~\mu$A \cite{WGK90a,RHK02}. 
During activation activation measurements (DC) 10$^9$~n/s at proton currents 
of $\approx~100~\mu$A are 
typically produced \\
\cite{RAH03}.

\begin{figure*}
\includegraphics[width=.95\textwidth]{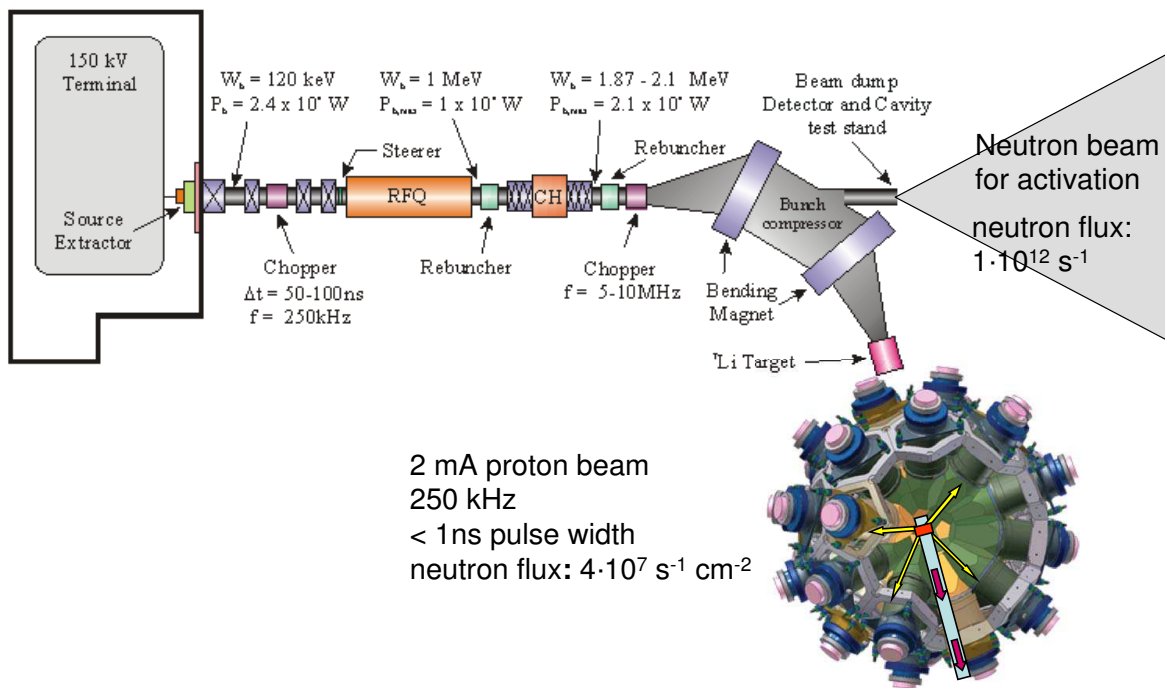}
\caption{Schematic layout of the Frankfurter Neutron Source FRANZ.}\label{franz}
\end{figure*}

The Stern-Gerlach-Zentrum SGZ recently founded at the University of Frankfurt 
allows to build and operate larger experiments now in accelerator physics, 
astrophysics and material science research. It was decided to develop an intense 
neutron generator within the next years. The proton driver LINAC consists of a 
high voltage terminal already under construction to provide primary proton beam 
energies of up to 150~keV. A volume type ion source will deliver a DC beam current 
of 100-250~mA at a proton fraction of 90\%. A low energy beam transport using 
four solenoids will inject the proton beam into a RFQ while a chopper at the 
entrance of the RFQ will create pulse lengths in the range of 100~ns at a 
repetition rate of up to 250~kHz. A drift tube cavity, witch delivers 
variable end energies between 1.9 and 2.1~MeV will be installed downstream of 
the RFQ. Finally a bunch compressor of the Mobley type forms a proton pulse 
length of 1 ns at the Li target. The maximum energies of the neutrons will 
be adjustable between $\approx$50~keV and $\approx$500~keV by the 
primary proton beam energy 
(see Figure~\ref{franz}).
 
\subsection{Time Of Flight Experiments}\label{ch_franz_tof}

In a first step, the use of the accelerator mentioned above and established 
lithium target technology, average beam currents of 0.1~mA are possible 
with a repetition rate between 250~kHz.

The second step would be to focus on improvements of the lithium target 
technology with the goal to increase the proton beam on target and hence 
the neutron flux. Improved cooling technologies allow targets with stable 
lithium layers that can handle up to 2~mA. This implies that without 
any major changes of the experimental setup compared to FZK (apart from the 
neutron production) an increase in neutron flux by a factor of 
1000 can be achieved.

\begin{figure}
\includegraphics[width=.45\textwidth]{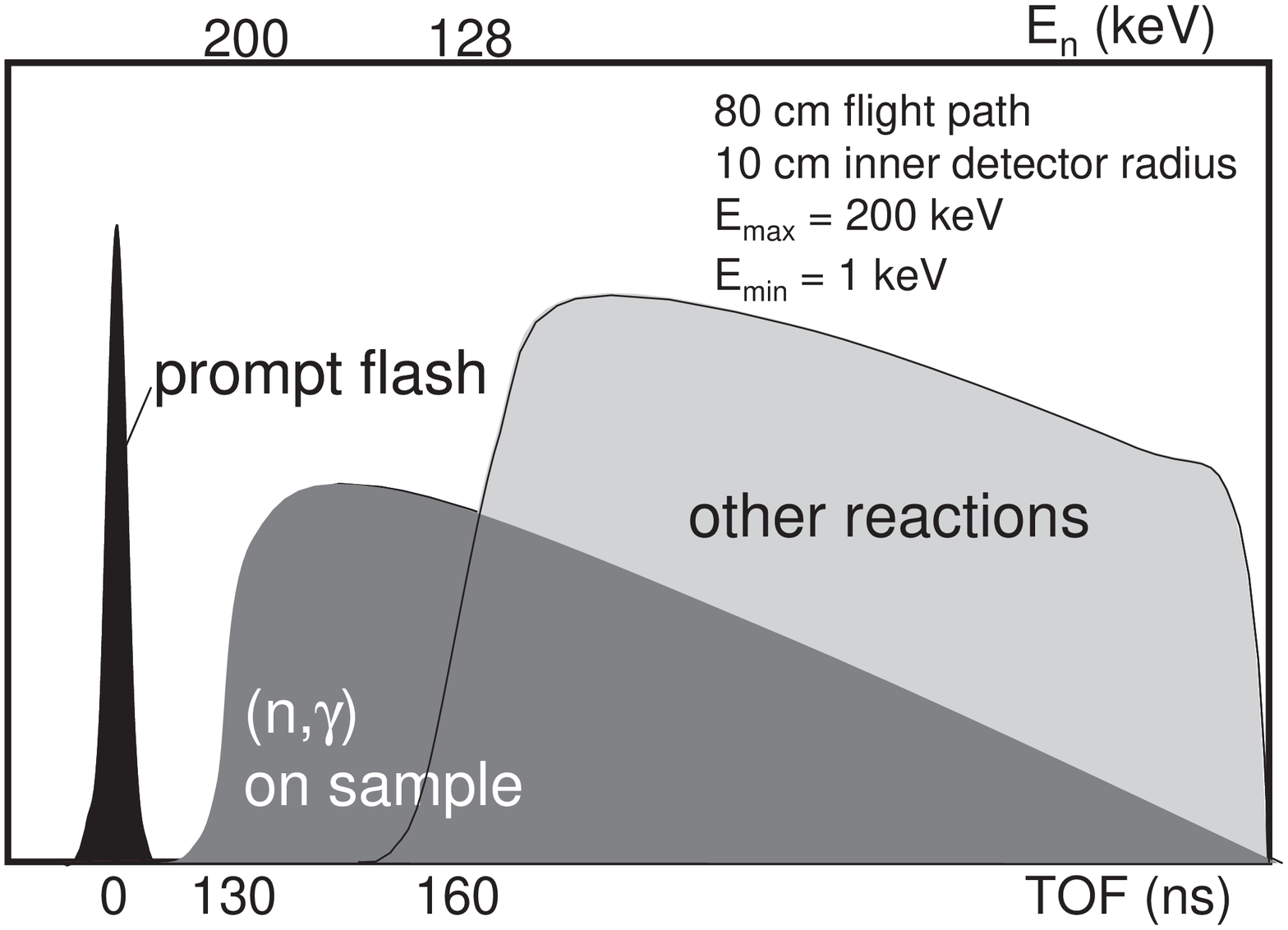}
\caption{Schematic TOF spectrum of the setup shown in Figure~\ref{franz} for a maximum neutron 
energy of 200~keV. The TOF region corresponding to neutron energies between 130 and 200~keV is
basically free of beam-related background.}\label{schematic_tof}
\end{figure}

\begin{figure}
\includegraphics[width=.45\textwidth]{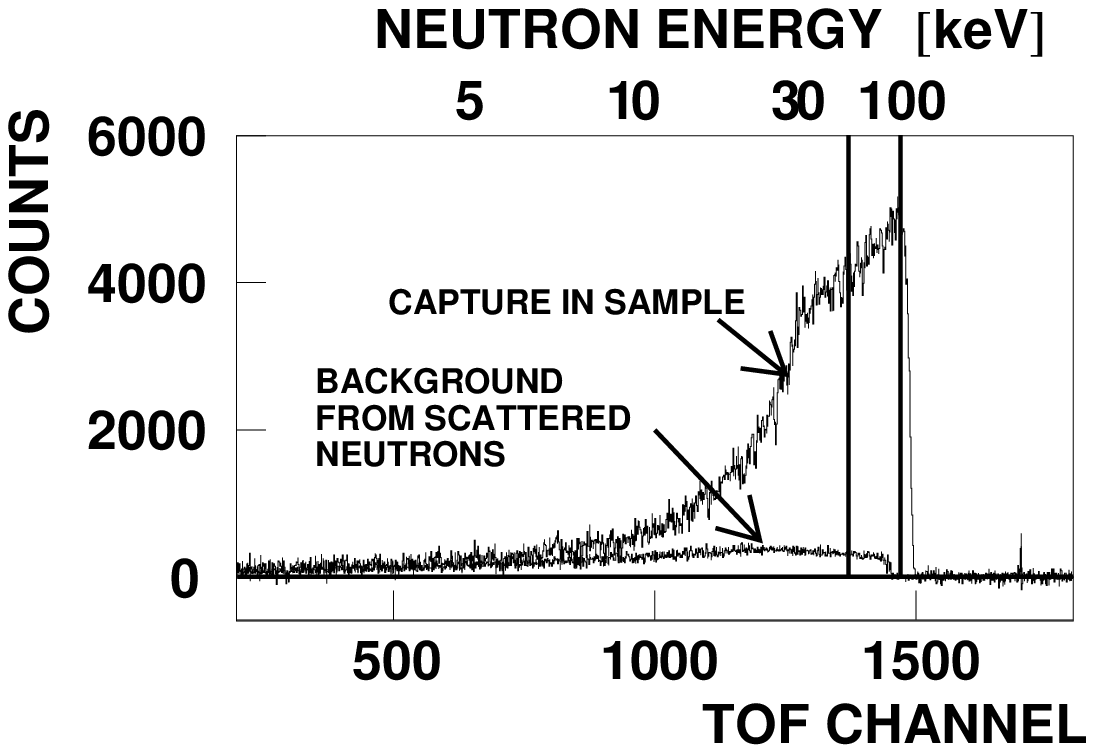}
\caption{TOF spectrum measured with the $^{129}$Xe for
100~keV maximum neutron energy. The background from
neutrons scattered at the sample is shown separately.}\label{129xe_tof}
\end{figure}

Important contributions to our understanding of the nucleosynthesis can then 
be made by measurements of extremely small capture cross sections on light elements, 
which are important since the respective isotopes are very abundant in the stars. 
Examples for this category of measurements are $^1$H\ncap,
which is important for Big Bang nucleosynthesis,
and $^{12}$C\ncap, $^{16}$O\ncap, $^{22}$Ne\ncap, which act as neutron poisons for the s-process.

The second category where significant contributions can be expected are ToF measurements 
on radioactive branch point nuclei. Some prominent examples of such measurements are
\ncap~experiments on $^{60}$Fe, $^{85}$Kr, $^{95}$Zr, $^{147}$Pm, $^{154}$Eu, 
$^{155}$Eu, $^{153}$Gd, and $^{185}$W \cite{CoR07}.

\begin{figure}
\includegraphics[width=.45\textwidth]{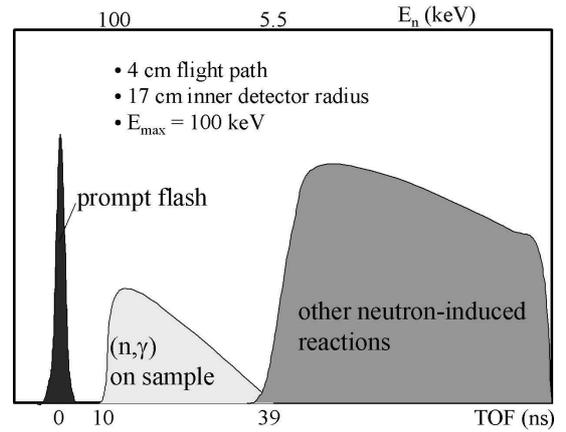}
\caption{Schematic TOF spectrum of the setup shown in Figure~\ref{franz} similar to 
the one shown in Figure~\ref{schematic_tof}. In this case a very short flight path
of only 4~cm was assumed. The intrinsic time resolution of the BaF$_2$ array and the proton
accelerator would be sufficient to disentangle neutron capture events from neutron-induced background 
as well as from the $\gamma$-flash based on their time relative to the proton pulse.}
\label{schematic_tof_short}
\end{figure}

The TOF experiments performed at the Research Center Karlsruhe had typically a flight path of $\approx80$~cm. 
The few exceptions, where a shorter flight path was used were carried out with 
Moxon Rae detectors 
\\ \cite{JaK96b}. 
This setup allowed extremely short flight paths of only 2~cm. For this particular
setup, the gain in neutron flux resulting from the short flight path was partly compensated by the significant loss
in $\gamma$-ray detection efficiency compared to the BaF$_2$ array. In principle both ideas -ultra-short flight path
and 4$\pi$ BaF$_2$ array- could be combined, but then the challenge is the enormous $\gamma$-flash immediately after the 
interaction of the proton pulse with the lithium target 
\\ \cite{RHH04,WHK06}. Figure~\ref{schematic_tof_short} shows an example of the time distribution of 
a possible realization. 
This approach is in particular interesting, 
if the activation method, which intrinsically has a much higher sensitivity, is not applicable. Prominent examples 
of such isotopes are
\ncap~experiments on $^{60}$Co, $^{65}$Zn, $^{86}$Rb, $^{89,90}$Sr, $^{127,127m}$Te, 
$^{134,137}$Cs
\\ \cite{CoR07}.     

\subsection{Activation experiments}\label{ch_franz_activation}

\begin{figure}[h]
\begin{center}
\includegraphics[width=.45\textwidth]{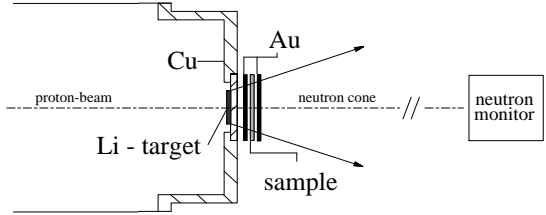}
\caption{Sketch of the activation setup at the Van de Graaff
accelerator.}\label{figactivation}
\end{center}
\end{figure}

While the detection efficiency of the (n,$\gamma$) products can be improved 
by up to one order of magnitude (the typical $\gamma$-ray efficiency of the currently 
mostly used 4$\pi$ Ge setup is 10\%), a real quantum leap is possible by 
improvements in the lithium target technology. Figure~\ref{figactivation} shows the typical activation
setup used at the Research Center Karlsruhe 
allowing a measurement relative to the well known $^{197}$Au(n,$\gamma$)$^{198}$Au 
cross section \cite{RaK88}.
Following the neutron irradiation, the activity of the gold foils can be determined via the 412~keV $\gamma$-ray
from the $^{198}$Au decay (t$_{1/2}$~=~2.69~d). 

\begin{figure}[h]
\begin{center}
\includegraphics[width=.45\textwidth]{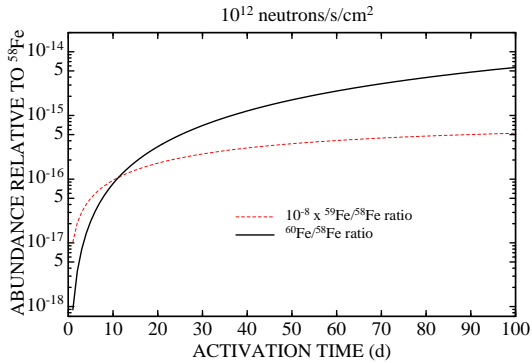}
\caption{Time dependence of the $^{59}$Fe and $^{60}$Fe content in a $^{58}$Fe sample during irradiation
with 10$^{12}$~n/s/cm$^2$ at FRANZ.}\label{figfe58_2n}
\end{center}
\end{figure}

Possible improvements of the target technique include 
rotating wheels as well as liquid lithium targets. If the target is able 
to handle the above mentioned $\approx$60~mA proton beam, neutron fluxes of 
about 10$^{12}$~n/s/cm$^2$ were available. Not only would that increase the 
sensitivity of the well established experimental techniques like cyclic 
activation \cite{Bee91} by three orders of magnitude, it would also allow to do activations 
via double neutron capture. This "advanced activation" would for instance 
allow the measurement of the unstable $^{59}$Fe (t$_{1/2}$ = 45~d) by activating a 
stable sample of $^{58}$Fe. The desired $^{59}$Fe\ncap~cross section could then be extracted by 
determining the $^{60}$Fe/$^{58}$Fe ratio via AMS \cite{KKF04}. This approach would provide a 
certain independence from the availability of radioactive samples (see Figure~\ref{figfe58_2n}).

\section{Summary}
Neutron-induced experiments have been successfully performed over the last 30 years at the Research Center Karlsruhe.
The basic ideas of the two well-established techniques, time of flight and activation, will be applied at the
FRANZ facility, which is currently under construction at the University of Frankfurt. Taking advantage of the 
latest developments in accelerator design, it will be possible to increase the sensitivity of both techniques
by about three orders of magnitude. This will open a new era in experimental nuclear astrophysics, since it will 
then be possible to investigate radioactive isotopes, which act as important branch points in the s-process, on a 
routinely basis.     

\section*{Acknowledgments} 
R.R. and R.P. are supported by the HGF Young Investigators Project VH-NG-327.

\newcommand{\noopsort}[1]{} \newcommand{\printfirst}[2]{#1}
  \newcommand{\singleletter}[1]{#1} \newcommand{\swithchargs}[2]{#2#1}
{}




\end{document}